\documentclass[aps,preprint,showpacs,floatfix]{revtex4}
\usepackage{epsfig}
\usepackage{amsmath,amsfonts,bm}
\begin{document}

\count255=\time\divide\count255 by 60 \xdef\hourmin{\number\count255}
  \multiply\count255 by-60\advance\count255 by\time
 \xdef\hourmin{\hourmin:\ifnum\count255<10 0\fi\the\count255}

\newcommand{\Dslash}{D\hspace{-0.7em}{ }\slash\hspace{0.2em}}
\newcommand\<{\langle}
\renewcommand\>{\rangle}
\renewcommand\d{\partial}
\newcommand\LambdaQCD{\Lambda_{\textrm{QCD}}}
\newcommand\tr{\mathop{\mathrm{Tr}}}
\newcommand\+{\dagger}
\newcommand\g{g_5}

\newcommand{\xbf}[1]{\mbox{\boldmath $ #1 $}}

\title{Optimal Parametrization of Deviations from Tribimaximal Form of
the Neutrino Mass Matrix}

\author{Christopher D. Carone}
\email{cdcaro@wm.edu}

\affiliation{Department of Physics, College of William \& Mary,
Williamsburg, VA 23187-8795, USA}

\author{Richard F. Lebed}
\email{Richard.Lebed@asu.edu}

\affiliation{Department of Physics, Arizona State University, Tempe,
AZ 85287-1504, USA}

\date{October, 2009}

\begin{abstract}
We obtain a general parametrization for the neutrino Majorana mass
matrix in which all possible deviations from tribimaximal
mixing are given by three complex parameters, and the deviation from
each tribimaximal mixing eigenvector is identified with precisely two
of them.  This parametrization provides a useful tool for classifying
the corrections to exact tribimaximal mixing in flavor models.
\end{abstract}


\pacs{14.60.Pq, 12.15.Ff, 11.30.Hv}
%

\maketitle

{\em Introduction}. Neutrino mass and mixing data to
date~\cite{Amsler:2008zzb} strongly favor a ``tribimaximal'' (TB)
mixing~\cite{TB} relationship between the neutrino mass and flavor
eigenstates. The unitary TB neutrino mixing matrix is given by
\begin{equation} \label{UTB}
U_{\rm TB} = \left(
\begin{array}{ccc}
 \sqrt{\frac{\textstyle 2}{\textstyle 3}} &
\frac{\textstyle 1}{\sqrt{\textstyle 3}} & 0 \\
-\frac{\textstyle 1}{\sqrt{\textstyle 6}} &
\frac{\textstyle 1}{\sqrt{\textstyle 3}} &
-\frac{\textstyle 1}{\sqrt{\textstyle 2}} \\
-\frac{\textstyle 1}{\sqrt{\textstyle 6}} &
\frac{\textstyle 1}{\sqrt{\textstyle 3}} &
+\frac{\textstyle 1}{\sqrt{\textstyle 2}}
\end{array}
\right) \, ,
\end{equation}
assuming a conventional choice of phases. $U_{\rm TB}$ is the special
case of the general (Pontecorvo)-Maki-Nakagawa-Sakata (MNS)~\cite{MNS}
form
\begin{equation} \label{angles}
U_{\rm MNS} = \left(
\begin{array}{ccc}
c_{13} c_{12} e^{i\beta_1} & c_{13} s_{12} e^{i\beta_2} & s_{13}
e^{-i\delta} \\
( -s_{12} c_{23} + s_{13} c_{12} s_{23} e^{i\delta} ) e^{i\beta_1} &
(  c_{12} c_{23} + s_{13} s_{12} s_{23} e^{i\delta} ) e^{i\beta_2} &
-c_{13} s_{23} \\
(  -s_{12} s_{23} - s_{13} c_{12} c_{23} e^{i\delta} ) e^{i\beta_1} &
(   c_{12} s_{23} - s_{13} s_{12} c_{23} e^{i\delta} ) e^{i\beta_2} &
\hphantom{+} c_{13} c_{23}
\end{array}
\right) \, ,
\end{equation}
where $s_{ij} \! \equiv \! \sin \theta_{ij}$, $c_{ij} \! \equiv \!
\cos \theta_{ij}$, for which $s_{13} \! = \! 0$ (and hence $\delta$ is
irrelevant), $s_{12} \! = \! 1/\sqrt{3}$ and $c_{12} \! = \!
\sqrt{2/3}$, $s_{23} \! = \! c_{23} \! = \! 1/\sqrt{2}$, and the
Majorana phases $\beta_{1,2}$, which do not contribute to neutrino
oscillation processes, are set to zero.  While deviations from TB
mixing can be described in terms of the angles of the MNS matrix, it
is more convenient to work directly with the eigenvectors of the
neutrino mass matrix ({\em i.e.}, the columns of $U_{\rm TB}$), which
do not depend on the conventions for parametrizing unitary matrices.

The purpose of this note is simply to show that deviations from the
tribimaximal form of the neutrino Majorana mass matrix are given {\em
uniquely} by a set of three independent complex parameters with the
property that setting any two of them to zero preserves one of the
vectors forming the columns of $U_{\rm TB}$.  Moreover, this property
remains true even when the nonzero parameter is arbitrarily
large. These special parameters are easily identified by the specific
form of the neutrino mass matrix in any given model.  In the physical
case, for which all the deviations from TB mixing are small, the
conventional angles and phases of the MNS matrix all have simple forms
at linear order in these parameters.  We compare our approach to other
relevant parametrizations at the end of the discussion.

{\em Diagonalization}. The MNS matrix brings the neutrino Majorana
mass matrix $M_{LL}$ to normal form, $diag(m_1 \, m_2 \, m_3) \! = \!
U_{\rm MNS}^T \, M_{LL} \, U^{\vphantom{T}}_{\rm MNS}$, with $m_i \!
\ge \! 0$.  In general, $M_{LL} \! \equiv \! M$ is a complex symmetric
matrix, meaning that its ``diagonalization'' by means of eigenvectors
is not completed by the usual means (as evidenced by the presence of
$U_{\rm MNS}^T$ rather than $U_{\rm MNS}^\dagger$), but the solution
procedure is well known~\cite{Takagi}: One solves
\begin{equation} \label{takagi_eq}
M v_i \! = \! m_i v_i^* \, ,
\end{equation}
where $m_i$ may always be chosen real and nonnegative, making the
normalized vectors $v_i$ unique up to overall signs.  In the
nondegenerate case, one may obtain an equivalent solution by
multiplying Eq.~(\ref{takagi_eq}) on the left by $M^*$:
\begin{equation} \label{takagi_eq2}
M^* \! M v_i \! = \! m_i^2 v_i \ .
\end{equation}
Here the $v_i$, which are now true normalized eigenvectors of $M^* \!
M$, are defined only up to overall phases; nevertheless, since the
neutrino mixing angles and Dirac phase $\delta$ are rephasing
invariant, solving Eq.~(\ref{takagi_eq2}) produces all observables
except the Majorana phases.  Noting that $M^* \! = \!  M^\dagger$ for
complex symmetric matrices, one sees that $\{m_i^2\}$ are simply the
modulus squares of the eigenvalues of $M$.

While the previous discussion is moot in the special case of exact TB
mixing (since $U_{\rm TB}$ is real), it is relevant for studying models
that deviate from the TB limit.  Using the explicit TB form
Eq.~(\ref{UTB}), it is easy to check that all diagonal matrices are in
a one-to-one correspondence with mass matrices $M$ of the form
\begin{equation}
M = \left(
\begin{array}{ccc}
x & y     & y     \\
y & x + v & y - v \\
y & y - v & x + v
\end{array}
\right) \, ,
\end{equation}
with complex eigenvalues $x-y$, $x+2y$, and
$x-y+2v$~\cite{Altarelli:2009kr}.  This form has three independent
complex degrees of freedom, $x$, $y$, and $v$, whereas the most
general 3$\times$3 complex symmetric matrix, has six parameters.  The
masses $m_{1,2,3}$ in normal form are the absolute values of these
eigenvalues; $M$ in this form produces Majorana phases in $U_{\rm
MNS}$, as discussed below.

The unique parametrization for the three remaining parameters, which
we call $\epsilon$, $\delta_1$, and $\delta_2$, such that introducing
each one of them in turn and keeping the other two zero precisely
preserves one of the vectors forming the columns of $U_{\rm TB}$, is
given by
\begin{equation} \label{master}
M = \left(
\begin{array}{ccc}
x              & y + \epsilon     & y + 2 \delta_1 \\
y + \epsilon   & x + v - \delta_1 & y - v + \delta_2 \\
y + 2 \delta_1 & y - v + \delta_2 & x + v + \epsilon + 3 \delta_1
\end{array}
\right) \, .
\end{equation}
The six parameters are given by
\begin{eqnarray}
x        & = & M_{11} \, , \nonumber \\
y        & = & +\frac 1 3 ( M_{12} + 2M_{13} + M_{22} - M_{33} ) \, ,
\nonumber \\
v        & = & -\frac 1 6 ( 6M_{11} + M_{12} - M_{13} - 5M_{22} -
M_{33} ) \, , \nonumber \\
\epsilon & = & +\frac 1 3 ( 2M_{12} - 2M_{13} - M_{22} + M_{33} ) \, ,
\nonumber \\
\delta_1 & = & -\frac 1 6 ( M_{12} - M_{13} + M_{22} - M_{33} ) \, ,
\nonumber \\
\delta_2 & = & -\frac 1 2 ( 2M_{11} + M_{12} + M_{13} - M_{22}
-2M_{23} - M_{33} ) \, . \label{invert}
\end{eqnarray}
This starting point, a fully general parametrization of $M_{LL}$ in
terms of all possible deviations from TB mixing, is also the starting
point in Ref.~\cite{Albright:2008rp}, which discusses the preservation
of TB eigenvectors.  Here, however, the preservation of TB
eigenvectors is used as an organizing principle in defining the
parametrization.  It is straightforward to check that the unique
choice of parameters $\delta_1$, $\epsilon$, and $\delta_2$ preserves
the vectors given by the first, second, and third columns of $U_{\rm
TB}$ in Eq.~(\ref{UTB}), respectively.  In other words, $\epsilon \!
\neq \!  0$ but $\delta_1 \! = \! \delta_2 \! = \! 0$ in this
parametrization preserves trimaximality, while $\delta_1$, $\delta_2$
parametrize all possible departures from it, and $\delta_2 \! \neq \!
0$ but $\epsilon \! = \delta_1 \! = \! 0$ ($\mu$-$\tau$ symmetry)
preserves bimaximality, while $\epsilon$ and $\delta_1$ parametrize
all possible departures from it. Introducing $\delta_1 \! \neq \! 0$
but $\epsilon \! = \! \delta_2 \! = \! 0$ preserves the remaining
orthogonal direction $(2,-1,-1)^T$.

Parametrizing $M$ as in Eq.~(\ref{master}) rather than $U_{\rm TB}$ is
more convenient in the construction of flavor models, which typically
provide textures of elements for Yukawa and Majorana matrices rather
than the matrices that diagonalize them.  Of course, $M$ here refers
to the Majorana mass matrix $M_{LL}$ in the mass basis for the charged
leptons (effectively, taking $Y_L$ diagonal).  In models that specify
nondiagonal forms for both $Y_L$ and $M_{LL}$, one faces the tricky
question of why the mismatch between the matrices that diagonalize
them nevertheless produces the TB angles to a high degree of accuracy.

The form of $M$ in Eq.~(\ref{master}) also allows for a
straightforward calculation of the neutrino masses and mixing angles
in a perturbative expansion simultaneously linear in $\epsilon$,
$\delta_1$, and $\delta_2$, which is appropriate due to the closeness
of the physical values to the TB solution.  Solving
Eq.~(\ref{takagi_eq2}) for the given $M$ produces the results
\begin{eqnarray} \label{masses}
m_1 & = & \left| x - y - \frac 1 2 \epsilon - \delta_1 +
\frac 1 3 \delta_2 \right| \, , \nonumber \\
m_2 & = & \left| x + 2y + \epsilon + 2\delta_1 + \frac 2 3 \delta_2
\right| \, ,
\nonumber \\
m_3 & = & \left| x - y + 2v + \frac 1 2 \epsilon + \delta_1 - \delta_2
\right| \, ,
\end{eqnarray}
and the corresponding modified eigenvectors
\begin{eqnarray}
v_1 = \frac{\textstyle 1}{\sqrt{\textstyle 6}} \left( \begin{array}{c}
2 + 2 \tilde \delta^*_2 \\
-1 - 3 \tilde \epsilon^* + 2 \tilde \delta^*_2 \\
-1 + 3 \tilde \epsilon^* + 2 \tilde \delta^*_2
\end{array} \right) \, , \nonumber \\
v_2 = \frac{\textstyle 1}{\sqrt{\textstyle 3}} \left( \begin{array}{c}
1 - 2 \tilde \delta_2 \\
1 + 3 \tilde \delta^*_1 + \tilde \delta_2 \\
1 - 3 \tilde \delta^*_1 + \tilde \delta_2
\end{array} \right) \, , \nonumber \\
v_3 = \frac{\textstyle 1}{\sqrt{\textstyle 2}} \left( \begin{array}{c}
-2 \tilde \epsilon + 2 \tilde \delta_1 \\
-1 + \tilde \epsilon + 2 \tilde \delta_1 \\
1 + \tilde \epsilon + 2 \tilde \delta_1
\end{array} \right) \, , \label{eigenvecs}
\end{eqnarray}
where
\begin{eqnarray} \label{new_params}
\tilde \epsilon & \equiv & \frac{1}{4} \frac{ {\rm Re} [ \epsilon
( x - y + v )^* ] - i \, {\rm Im} ( \epsilon \, v^* ) }{ {\rm Re} [ v
( x - y + v )^* ] } \, , \nonumber \\
\tilde \delta_1 & \equiv & - \frac { {\rm Re} [ \delta_1 ( 2x + y + 2v
)^* ] + i \, {\rm Im} [ \delta_1 ( 3y - 2v)^* ] }{ {\rm Re} [ ( 3y -
2v ) ( 2x + y + 2v )^* ] } \, , \nonumber \\
\tilde \delta_2 & \equiv & \frac{1}{9} \frac{ {\rm Re} [ \delta_2
( 2x + y)^* ] - 3i \, {\rm Im} ( \delta_2 y^* ) }{ {\rm Re} [ y ( 2x +
y )^* ] } \, .
\end{eqnarray}
In models in which all the parameters are real, the combinations in
Eq.~(\ref{new_params}) simplify to $\tilde \epsilon \! = \!
\epsilon/4v$, $\tilde \delta_1 \! = \!  -\delta_1/(3y - 2v)$, and
$\tilde \delta \! = \! \delta_2 / 9y$.  In this case, the independence
of the eigenvectors on $x$ follows simply from the fact that it enters
$M$ as a multiple of the identity matrix.

These expressions also elucidate the precise conditions for small
departures from the TB form: $|\tilde \epsilon|$, $|\tilde \delta_1
|$, $|\tilde \delta_2 | \! \ll \! 1$.  One can easily check that the
pathological cases, {\it e.g.}, $v \! = \! 0$ but $\epsilon \! \neq \!
0$, lead to degenerate masses, and $U_{\rm MNS}$ is no longer
guaranteed to assume a TB form at leading order.

Now, taking into account the overall phase of $v_i$ appearing in the
solution of Eq.~(\ref{takagi_eq}) but not Eq.~(\ref{takagi_eq2}), one
assembles the new MNS matrix
\begin{equation} \label{our_MNS}
U_{\rm MNS} = \left( v_1 \left| \, v_2 \, \right| v_3 \right) P \, ,
\end{equation}
where $P \! \equiv \! diag\{ \exp( -\frac i 2 {\rm Arg} \, \mu_i )
\}$, with $\mu_i$ being the complex mass parameters in
Eq.~(\ref{masses}): $m_i \! \equiv \! |\mu_i|$.  $U_{\rm MNS}$ reduces
to TB form with nontrivial Majorana phases when $\tilde \epsilon$,
$\tilde \delta_1$, and $\tilde \delta_2$ are set to zero; however, one
also notes a different phase convention from Eq.~(\ref{angles}), in
which $U_{{\rm MNS}, \, 23}$ and $U_{{\rm MNS}, \, 33}$ are manifestly
real.  Multiplication of $U_{\rm MNS}$ on the left by an arbitrary
diagonal phase matrix ({\it i.e.}, rephasing its rows $U_{{\rm MNS},
\, ij} \! \to \! e^{i\theta_i} U_{{\rm MNS}, \, ij}$), which changes
none of the observables, resolves this discrepancy.  To first order in
$\tilde \epsilon$, $\tilde \delta_1$, and $\tilde
\delta_2$, one obtains the Euler angles
\begin{eqnarray} \label{angle_pred}
s_{13} & = & \sqrt{2} | \tilde \delta_1 - \tilde \epsilon | \, ,
\nonumber \\
s_{23} & = & \frac{1}{\sqrt{2}} | 1 - \tilde \epsilon - 2 \tilde
\delta_1 | \simeq \frac{1}{\sqrt{2}} [ 1 - {\rm Re} \, ( \tilde
\epsilon + 2 \tilde \delta_1 ) ] \, , \nonumber
\\
s_{12} & = & \frac{1}{\sqrt{3}} | 1 - 2 \tilde \delta_2 | \simeq
\frac{1}{\sqrt{3}} ( 1 - 2 {\rm Re} \, \tilde \delta_2 ) \, .
\end{eqnarray}
As expected, if bimaximality is preserved ($\epsilon \! = \! \delta_1
\! = \! 0$), then $s_{13} \! = \! 0$, and the atmospheric
angle value is given by $s_{23} \! = \! 1/\sqrt{2}$ exactly.  One
also notes that $\delta_2$ alone in this approximation parametrizes
departures from the solar angle value $s_{12} \! = \! 1/\sqrt{3}$.
Using global fit values of the neutrino mixing
angles~\cite{Schwetz:2008er},
\begin{eqnarray}
\sin^2 \theta_{13} & = & 0.01^{+0.016}_{-0.011} \, , \nonumber \\
\sin^2 \theta_{23} & = & 0.50^{+0.07}_{-0.06} \, , \nonumber \\
\sin^2 \theta_{12} & = & 0.304^{+0.022}_{-0.016} \, ,
\end{eqnarray}
one finds that the parameters are indeed small:
\begin{eqnarray}
| \tilde \delta_1 - \tilde \epsilon | & = & 0.07^{+0.06}_{-0.04} \, ,
\nonumber \\
{\rm Re} \, ( \tilde \epsilon + 2 \tilde \delta_1 ) & = &
0.00^{+0.06}_{-0.07} \, ,
\nonumber \\
{\rm Re} \, \tilde \delta_2 & = & 0.023^{+0.013}_{-0.017} \, .
\end{eqnarray}

The rephasing that leads to real $U_{{\rm MNS}, \, 23}$ and $U_{{\rm
MNS}, \, 33}$  is
\begin{equation}
\theta_i = \frac 1 2 {\rm Arg} ( \mu_3 ) + \eta_i \, , \ {\rm where} \
\eta_{1,2,3} = 0, \, +{\rm Im} ( \tilde \epsilon + 2 \tilde \delta_1 )
, \, -{\rm Im} ( \tilde \epsilon + 2 \tilde \delta_1 ) \, .
\end{equation}
The MNS phases may then be obtained from
Eqs.~(\ref{eigenvecs})--(\ref{our_MNS}):
\begin{eqnarray}
\delta & = & - {\rm Arg} ( \tilde \delta_1 - \tilde \epsilon ) \, ,
\nonumber \\
\beta_1 & = & \frac 1 2 \, ( {\rm Arg} \, \mu_3 - {\rm Arg} \, \mu_1 )
- {\rm Im} \, \tilde \delta_2 \nonumber \, , \\
\beta_2 & = & \frac 1 2 \, ( {\rm Arg} \, \mu_3 - {\rm Arg} \, \mu_2 )
- 2 \, {\rm Im} \, \tilde \delta_2 \, ,
\end{eqnarray}
and the Jarlskog parameter is given by
\begin{equation}
J = \frac 1 3 \, {\rm Im} \, ( \tilde \delta_1 - \tilde \epsilon ) \, .
\end{equation}
Note that $\beta_1$, $\beta_2$, and $\delta$ can all be nonzero even
when $\tilde \epsilon$, $\tilde \delta_1$, $\tilde \delta_2 \! \to \!
0$.  Although the mass-squared splitting ratio $\Delta m_{21}^2 /
\Delta m_{32}^2 = 0.0319^{+0.0028}_{-0.0018}$ is experimentally
small~\cite{Schwetz:2008er}, it is nonvanishing in the TB limit:
\begin{equation}
\frac{\Delta m_{21}^2}{\Delta m_{32}^2} =
- \frac {3 {\rm Re} \, \left[ y ( 2x + y )^* \right] } { {\rm Re} \,
\left[ (3y - 2v) (2x + y + 2v )^* \right] }
+ O ( \tilde \epsilon, \tilde \delta_1, \tilde \delta_2 ) \, .
\end{equation}

{\em Discussion}. A number of parametrizations of the MNS matrix have
appeared in the literature~\cite{others}.  Two recent parametrizations
in particular describe deviations from TB in terms of three (real)
parameters plus the CP-violating phase $\delta$.  The relations of
these two sets, $r$, $s$, and $a$ in Ref.~\cite{King:2007pr} and
$\epsilon_{13}$, $\epsilon_{21}$, and $\epsilon_{32}$ in
Ref.~\cite{Pakvasa:2007zj}, to those defined here at linear order are
given by
\begin{eqnarray}
| \tilde \delta_1 - \tilde \epsilon | & = & \frac r 2 =
\frac{\epsilon_{13}}{\sqrt{2}} \, , \nonumber \\
{\rm Re} \, ( \tilde \epsilon + 2 \tilde \delta_1 ) & = & -a =
-\epsilon_{32} \, , \nonumber \\
{\rm Re} \, \tilde \delta_2 & = & -\frac s 2 =
-\frac{\epsilon_{21}}{\sqrt{2}} \, .
\end{eqnarray}
The advantages of the current parametrization have already been
identified, but to summarize: (1) The parameters $\tilde \epsilon$,
$\tilde \delta_1$, $\tilde \delta_2$ have a natural interpretation in
terms of which TB eigenvector each preserves, even when the chosen
parameter is not perturbatively small. (2) The mixing angles and Dirac
phase of the MNS matrix have simple expressions at linear order in
$\tilde \epsilon$, $\tilde \delta_1$, $\tilde \delta_2$.  (3) Our
parameters allow a direct analysis of mass matrix corrections that
lead to a deviation from exact TB mixing in flavor models.  For
example, in one well-known model~\cite{verydiff}, the entries of
$M_{LL}$ are determined by four vacuum expectation values (vevs) of
Higgs triplets, labeled $a$, $b$, $c$, $d$.  From the mass matrix
given in Eq.~(19) of Ref.~\cite{verydiff}, it is straightforward to
check [using, for example, Eq.~(\ref{invert})] that
\begin{equation}
\delta_1=\delta_2=0 \,\,\,\,\, \mbox{ and } \,\,\,\,\, \epsilon=b-c
\,\,\,.
\end{equation}
Our parametrization shows immediately that this model is always
exactly trimaximal.  The deviation from bimaximality is reflected in
$s_{13} \! = \! \frac{1}{2\sqrt{2}} \left| \frac{b-c}{c-a} \right|$
when $|b-c| \! \ll \! |c-a|$ and all vevs are real.

\vspace{-0.5em}
\section*{Acknowledgments}
\vspace{-0.5em}
This work was supported by the NSF under Grant Nos.\ PHY-0757481 (CDC)
and PHY-0757394 (RFL).

\end{document}